\documentclass[12pt]{article}

\newcommand{\firstG}{\mathcal{G}_1}
\newcommand{\secondG}{\mathcal{G}_2}
\newcommand{\thirdG}{\mathcal{G}_3}
\newcommand{\fourthG}{\mathcal{G}_4}
\newcommand{\fifthG}{\mathcal{G}_5}
\newcommand{\sixthG}{\mathcal{G}_6}

\title{\bf Determining the shape of the Universe 
using discrete sources}

\author{
G.I. Gomero\thanks{german@ift.unesp.br}, \\
\\
Instituto de F\'{\i}sica Te\'orica, \\
Universidade Estadual Paulista,  \\
Rua Pamplona, 145 \\
S\~ao Paulo, SP 01405--900, Brazil
}

\begin{document}

\date{\today}

\maketitle

\begin{abstract} \noindent
Suppose we have identif\/ied three clusters of galaxies as
being topological copies of the same object. How does this
information constrain the possible models for the shape
of our Universe? It is shown here that, if our Universe has
f\/lat spatial sections, these multiple images can be
accommodated within any of the six classes of compact
orientable 3-dimensional f\/lat space forms. Moreover, the
discovery of two more triples of multiple images in the
neighbourhood of the f\/irst one, would allow the
determination of the topology of the Universe, and in most
cases the determination of its size. 
\end{abstract}

\section{Introduction}

The last two decades have seen a continuously increasing 
interest in studying cosmological models with multiply 
connected spatial sections (see~\cite{Review} and references 
therein). Since observational cosmology is becoming an 
increasingly high precision science, it would be of wide 
interest to develop methods to systematically construct 
specif\/ic candidates for the shape of our Universe in order 
to analyse whether these models are consistent with 
observational data.

Since one of the simplest predictions of cosmological models 
with multiply connected spatial sections is the existence of 
multiple images of discrete cosmic objects, such as clusters 
of galaxies,%
\footnote{Provided that the scale of compactif\/ication is 
small enough (see~\cite{Detect} and \cite{GR02}).}
the following question immediately arises: Suppose we have 
identif\/ied three clusters of galaxies as being dif\/ferent 
topological copies of the same object, how does this information
constrain the possible models for the shape of our Universe?
The initial motivation for this work was the suggestion of
Roukema and Edge that the X--ray clusters RXJ 1347.5--1145 and
CL 09104+4109 may be topological images of the Coma cluster
\cite{RE97}. Even if these particular clusters turn out not to
be topological copies of the same object, the suggestion of
Roukema and Edge raises an interesting challenge. \emph{What
if} one day a clever astrophysicist discovers three topological
copies of the same object?

It is shown here that these (would be) multiple images could
be accommodated within any of the six classes of compact
orientable 3-dimensional f\/lat space forms. Moreover, 
and this is the main result of this paper, the discovery 
of two more triples of multiple images in the neighbourhood 
of the f\/irst one, would be enough to determine the 
topology of the Universe, and in most cases even its size. 
Thus, two interesting problems appear now, (i) does our 
present knowledge of the physics of clusters of galaxies 
(or alternatively, of quasars) may allow the identif\/ication 
of a triple of multiple images if they actually exist?, and 
(ii) given that such an identif\/ication has been achieved, 
how easy can other triples of topological copies near the 
f\/irst one be identif\/ied? The present paper does not 
deal with these two problems, however it should be noticed 
that a recent method proposed by A. Bernui and me in 
\cite{BerGo} (see also \cite{Gomero}) could be used to test, 
in a purely geometrical way, the hypothesis that any two 
given clusters of galaxies are topological copies.

The model building procedure is explained in the next section, 
while section \ref{Examples} presents some numerical examples 
illustrating specif\/ic candidates for the shape of our 
Universe, under the pressumed validity of the Roukema--Edge 
hypothesis. In section \ref{Decide} it is discussed the main 
result of this paper: how the topology of space could be 
determined with the observation of just two more triples of 
images; and how, in most cases, one could even determine the 
size of our Universe. Finally, section \ref{Concl} consists 
of discussions of the results presented in this letter and 
suggestions for further research.

\section{Model Building} 
\label{ModBuild}

Suppose that three topological copies of the same cluster 
of galaxies have been identif\/ied. Let $C_0$ be the nearest 
copy from us, $C_1$ and $C_2$ the two other copies, $d_1$ 
and $d_2$ the distances from $C_0$ to $C_1$ and $C_2$ 
respectively, and $\theta$ the angle between the geodesic 
segments $\overline{C_0C_1}$ and $\overline{C_0C_2}$. 
Roukema and Edge \cite{RE97} have suggested an example of 
this conf\/iguration, the Coma cluster being $C_0$ and the 
clusters RXJ 1347.5--1145 and CL 09104+4109 being $C_1$ and 
$C_2$ (or vice versa). The distances of these clusters to 
Coma are 970 and 960$h^{-1}$ $Mpc$ respectively (for 
$\Omega_0=1$ and $\Lambda=0$), and the angle between them, 
with the Coma cluster at the vertex, is $\approx \! 88^o$. 
Under the assumption that these multiplicity of images were 
due to two translations of equal length and in orthogonal 
directions, they constructed FL cosmological models whose 
compact f\/lat spatial sections of constant time were (i) 
3-torii, (ii) manifolds of class $\secondG$, or (iii) 
manifolds of class $\fourthG$, all of them with square
cross sections, and scale along the third direction larger
than the depth of the catalogue of X-ray clusters used in
the analysis.

Let us consider the possibility that at least one of the 
clusters $C_i$ is an image of $C_0$ by a screw motion, 
and do not assume that the distances from $C_0$ to $C_1$ 
and $C_2$ are equal, nor that they form a right angle 
(with $C_0$ at the vertex). It is shown in this section 
that one can accommodate this generic conf\/iguration of clusters
within any of the six classes of compact orientable 
3-dimensional f\/lat space forms, thus providing a 
plethora of models for the shape of our Universe consistent 
with the (would be) observational fact that these clusters 
are in fact the same cluster. Moreover, one could also 
consider the possibility that one of the clusters $C_i$ 
is an image of $C_0$ by a glide ref\/lection, thus giving 
rise to non--orientable manifolds as models for the shape 
of space. However, these cases will not be considered here 
since they do not give qualitatively dif\/ferent results, 
and the corresponding calculations can be done whenever 
needed.

The dif\/feomorphic and isometric classif\/ications of 
3-dimensional Euclidean space forms given by Wolf in 
\cite{Wolf} were described in detail by Gomero and Rebou\c{c}as 
in \cite{GR02}. The generators of the six dif\/feomorphic 
compact orientable classes are given in Table \ref{Tb:OESF}, 
where an isometry in Euclidean 3-space is denoted by $(A,a)$,
$a$ is a vector and $A$ is an orthogonal transformation, and 
the action is given by 
\begin{equation}
\label{action}
(A,a) : p \mapsto Ap + a \; ,
\end{equation}
for any point $p$. The orientation preserving orthogonal 
transformations that appear in the classif\/ication of the 
Euclidean space forms take the matrix forms 
\begin{eqnarray}                                         
\label{Rot3}
A_1 = \left( \begin{array}{ccc}
               1 &  0 & 0 \\
               0 & -1 & 0 \\
               0 &  0 & -1
      \end{array} \right) \; , & 
A_2 = \left( \begin{array}{ccc}
              -1 & 0 & 0 \\
               0 & 1 & 0 \\
               0 & 0 & -1
      \end{array} \right) \; , & 
A_3 = \left( \begin{array}{ccc}
              -1 &  0 & 0 \\
               0 & -1 & 0 \\
               0 &  0 & 1
      \end{array} \right) \; , \nonumber \\ \\
B = \left( \begin{array}{ccc}
              1 & 0 & 0 \\
              0 & 0 & -1 \\
              0 & 1 & -1
    \end{array} \right) \; , & 
C = \left( \begin{array}{ccc}
              1 & 0 & 0 \\
              0 & 0 & -1 \\
              0 & 1 & 0
    \end{array} \right) \quad\mbox{and} & 
D = \left( \begin{array}{ccc}
              1 & 0 & 0 \\
              0 & 0 & -1 \\
              0 & 1 & 1
    \end{array} \right) \; , \nonumber
\end{eqnarray}
in the basis formed by the set $\{a,b,c\}$ of linearly 
independent vectors that appear in Table~1. We will f\/it the
set of multiple images $\{C_0,C_1,C_2\}$ within manifolds of 
classes $\secondG-\sixthG$, since the class $\firstG$ (the 
3--torus) is trivial. 

\begin{table}[t]
\begin{center}
\begin{tabular}{|c|*{6}{|c}|} \hline
Class & $\firstG$ & $\secondG$ & $\thirdG$ & $\fourthG$ & 
$\fifthG$ & $\sixthG$ \\ \hline 
& & & & & & ($A_1$,$a$) \\
Generators & $a$, $b$, $c$ & ($A_1$,$a$), $b$, $c$ & 
($B$,$a$), $b$, $c$ & ($C$,$a$), $b$, $c$ & 
($D$,$a$), $b$, $c$ & ($A_2$,$b+c$) \\
& & & & & & ($A_2$,$b-c$) \\ \hline
\end{tabular}   
\caption[Compact orientable $3$-dimensional Euclidean space 
forms.] {\label{Tb:OESF} \footnotesize Dif\/feomorphism classes 
of compact orientable $3$-dimensional Euclidean space forms. 
The f\/irst row contains Wolf's notation for each class, and the 
second gives the generators of the corresponding covering 
groups.}
\end{center}
\end{table}

Let us f\/irst deal with the classes $\secondG-\fifthG$. The 
generators for the corresponding covering groups are $\alpha 
= (A,a)$, $\beta=(I,b)$ and $\gamma=(I,c)$, with $A=A_1,B,C$ 
and $D$ for the classes $\secondG$, $\thirdG$, $\fourthG$ and 
$\fifthG$ respectively, and $I$ is the identity transformation. 
For these classes we will consider the following non-trivial 
conf\/iguration: denoting the position of $C_0$ by $p$, $C_1$ 
is located at $\alpha(p)$ and $C_2$ at $\beta(p)$. The 
conf\/iguration in which $C_2$ is located at $\gamma(p)$ is 
equivalent to the former, while the conf\/iguration in which 
$C_1$ and $C_2$ are images of $C_0$ by pure translations 
(strictly possible only in $\secondG$, and a convenient 
approximation in $\fourthG$ if $\theta \approx 90^o$, and the 
distances of $C_1$ and $C_2$ to $C_0$ are almost equal, as is 
the case in the Roukema--Edge hypothesis) is equivalent to 
that of a torus.

For space forms of the classes $\secondG-\fifthG$ the following 
facts are easily derivable from the generators of their 
corresponding covering groups (see \cite{GR02} for details):
\begin{enumerate}
\item The vector $a$ is orthogonal to both $b$ and $c$.
\item \label{angle} The angle between $b$ and $c$ is a free 
parameter 
for the class $\secondG$, while its value is f\/ixed to be 
$120^o$, $90^o$ and $60^o$ for the classes $\thirdG$, $\fourthG$ 
and $\fifthG$ respectively.
\item Denoting by $|a|$ the length of the vector $a$, and 
similarly for any other vector, one has that $|b| = |c|$ for the 
classes $\thirdG-\fifthG$, while both lengths are independent 
free parameters in the class $\secondG$. Moreover, in all classes 
$\secondG-\fifthG$, $|a|$ is an independent free parameter.
\item Denoting the canonical unitary basis vectors in Euclidean 
space by $\{\hat{\imath},\hat{\jmath},\hat{k}\}$, one can
always write $a = |a| \hat{\imath}$, $b = |b| \hat{\jmath}$ and 
$c = |c| \cos \varphi \hat{\jmath} + |c| \sin \varphi \hat{k}$, 
for the basis $\{a,b,c\}$, where $\varphi$ is the angle between 
$b$ and $c$, as established in the item \ref{angle}.
\end{enumerate}

\begin{table}[t]
\begin{center}
\begin{tabular}{*{4}{|c}|} \hline & & & \\
Class & $\alpha(p)$ & $\delta_{\alpha}(p)$ & $\delta_{\alpha}(p) 
\cos(\alpha,\beta)$ \\ & & & \\ \hline \hline
& & & \\ 
$\secondG$ & $(x+|a|,-y, -z)$ 
& $\sqrt{|a|^2+4(y^2+z^2)}$ & $-2y$ 
\\ & & & \\ \hline 
& & & \\ 
$\thirdG$ & $(x+|a|, -\frac{1}{2} y - \! \frac{\sqrt{3}}{2} z, 
\, \frac{\sqrt{3}}{2} y - \! \frac{1}{2} z)$ 
& $\sqrt{|a|^2 + 3(y^2+z^2)}$ & $- \frac{\sqrt{3}}{2}(\sqrt{3}y 
+ z)$ \\ 
& & & \\ \hline
& & & \\ 
$\fourthG$ & $(x+|a|,-z,y)$ 
& $\sqrt{|a|^2+2(y^2+z^2)}$ & $-(y+z)$ \\ 
& & & \\ \hline
& & & \\ 
$\fifthG$ & $(x+|a|, \frac{1}{2} y - \! \frac{\sqrt{3}}{2} z, \, 
\frac{\sqrt{3}}{2} y + \frac{1}{2} z)$ 
& $\sqrt{|a|^2+(y^2+z^2)}$ & $-\frac{1}{2}(y + \sqrt{3}z)$ 
\\ & & & \\ \hline
\end{tabular}
\caption[Position and metrical expressions for $C_1$.] 
{\label{Tb:Express} \footnotesize The second column gives the 
position of $C_1$ for each class of the manifolds considered in 
the f\/irst column. The third column gives the distance between 
$C_0$ and $C_1$, and the last one the cosine of the angle 
between the segments $\overline{C_0C_1}$ and $\overline{C_0C_2}$, 
$\cos(\alpha,\beta)$.}
\end{center}
\end{table}

Writing $p=(x,y,z)$ for the components of the position of $C_0$ 
in the basis $\{\hat{\imath},\hat{\jmath},\hat{k}\}$,%
\footnote{Note that the origin of a coordinate system is 
implicitly determined by the axes of rotation of the orthogonal 
transformations in (\ref{Rot3}), and can be taken as the centre 
of the fundamental polyhedron for the corresponding manifold. 
Moreover, this origin does not necessarily coincide with the 
position of our galaxy.} 
one can easily work out the expressions for the components 
of the position of $C_1$, $\alpha(p)$, the distance function 
$\delta_{\alpha}(p)$, and the cosine of the angle between 
$\overline{C_0C_1}$ and $\overline{C_0C_2}$, $\cos(\alpha,\beta)$. 
The resulting expressions are shown in Table~\ref{Tb:Express}.

For the conf\/iguration we are dealing with, one trivially has 
$d_2=\delta_{\beta}(p)=|b|$, since $\beta$ is a pure translation. 
More interestingly, from $\delta_{\alpha}(p) = d_1$ and 
$\cos(\alpha,\beta)=\cos\theta$, one can partially solve the 
equations for the components of the position of $C_0$. The 
resulting expressions are shown in Table~\ref{Tb:Positions}. 
Observe that for each class we have two solutions in terms of the 
free parameter $|a|$. For the classes $\thirdG-\fifthG$ the two 
solutions are those for which $d_1 \!\cos\theta$ is given by the 
fourth column in Table~\ref{Tb:Express}. 

Two remarks are in order here. First, it is convenient to write 
down the components of the position of $C_0$ in terms of the 
parameter $|a|$, because this parameter can be easily determined 
once two more triples of multiple images, say $\{D_0,D_1,D_2\}$ 
and $\{E_0,E_1,E_2\}$, in the neighbourhood of $\{C_0,C_1,C_2\}$ 
have been identif\/ied, as shown in Section~\ref{Decide}.%
\footnote{Actually, it can be done much more than that. If the 
topology of the Universe turns out to be of any of the classes 
$\secondG-\sixthG$, the triples $\{D_0,D_1,D_2\}$ and 
$\{E_0,E_1,E_2\}$ would be enough to decide which topology our 
Universe has, and except in the case of $\secondG$ and a 
conf\/iguration in $\sixthG$, it would be possible to specify 
completely the parameters of the manifold that models the spatial 
sections of the spacetime.}
Once this has been done, the positions of $C_0$, $D_0$ and $E_0$ 
can be used to predict multiple images of them due to the inverse 
isometry $\alpha^{-1}$, thus yielding a def\/initive observational 
test for the hypothesis of the multiply connectedness of our 
Universe. Second, note that the $x-\,$coordinate is not 
constrained by this conf\/iguration of topological images. This 
freedom of the $x-\,$coordinate is a consequence of homogeneity 
of manifolds of classes $\secondG-\fifthG$ along the $x-\,$axis. 
This \emph{partial} homogeneity is due to the fact that the 
orthogonal transformations involved in the corresponding covering 
groups have the $x-\,$axis as their axis of rotation.

\begin{table}[t]
\begin{center}
\begin{tabular}{*{3}{|c}|} \hline & & \\
Class & $y$ & $z$ \\ & & \\ \hline \hline
& & \\ 
$\secondG$ & $-\frac{1}{2} \, d_1 \cos\theta$ & $\pm \frac{1}{2} 
\sqrt{d_1^2 \sin^2 \theta - |a|^2}$ \\ & & \\ \hline 
& & \\ 
$\thirdG$ & $\pm \frac{\sqrt{3}}{6} \sqrt{d_1^2 \sin^2 \theta - 
|a|^2} - \frac{1}{2} \, d_1\cos\theta$ & $\mp \frac{1}{2} 
\sqrt{d_1^2 \sin^2 \theta - |a|^2} - \frac{\sqrt{3}}{6} \, 
d_1\cos\theta$ \\ & & \\ \hline
& & \\ 
$\fourthG$ & $\pm \frac{1}{2} \sqrt{d_1^2 \sin^2 \theta - |a|^2} - 
\frac{1}{2} \, d_1\cos\theta$ & $\mp \frac{1}{2} \sqrt{d_1^2 
\sin^2 \theta - |a|^2} - \frac{1}{2} \, d_1\cos\theta$ \\ & & \\ 
\hline
& & \\ 
$\fifthG$ & $\pm \frac{\sqrt{3}}{2} \sqrt{d_1^2 \sin^2 \theta - 
|a|^2} - \frac{1}{2} \, d_1\cos\theta$ & $\mp \frac{1}{2} 
\sqrt{d_1^2 \sin^2 \theta - |a|^2} - \frac{\sqrt{3}}{2} \, 
d_1\cos\theta$ \\ & & \\ \hline
\end{tabular}
\caption[Position and metrical expressions for $C_1$.] 
{\label{Tb:Positions} \footnotesize Partial solutions for the 
positions of $C_0$ for f\/lat manifolds of the classes 
$\secondG$--$\fifthG$.}
\end{center}
\end{table}

We now f\/it the multiple images $\{C_0,C_1,C_2\}$ within 
manifolds of class $\sixthG$. The generators for the covering 
group of a manifold of this class are $\alpha=(A_1,a)$, 
$\beta=(A_2,b+c)$ and $\mu=(A_2,b-c)$. The vectors $\{a,b,c\}$ 
are mutually orthogonal but their lengths are free parameters. 
For manifolds of class $\sixthG$ we have two possible 
conf\/igurations, both of them with $C_0$ located at $p$,
\begin{enumerate}
\item $C_1$ located at $\alpha(p)$ and $C_2$ at $\beta(p)$, and 
\item $C_1$ located at $\beta(p)$ and $C_2$ at $\mu(p)$.
\end{enumerate}
The case in which $C_1$ is at $\alpha(p)$ and $C_2$ at $\mu(p)$ 
is equivalent to the f\/irst conf\/iguration.

The expressions for the distances $\delta_{\alpha}(p)$, 
$\delta_{\beta}(p)$ and $\delta_{\mu}(p)$, and angles 
$\cos(\alpha,\beta)$ and $\cos(\beta,\mu)$ are
\begin{eqnarray}
\delta_{\alpha}(p) & = & \sqrt{|a|^2 + 4(y^2+z^2)} \nonumber \\
\label{DistFunc}
 \delta_{\beta}(p) & = & \sqrt{|b|^2 + 4x^2 + (2z-|c|)^2} \\
   \delta_{\mu}(p) & = & \sqrt{|b|^2 + 4x^2 + (2z+|c|)^2} 
\nonumber
\end{eqnarray}
and
\begin{eqnarray}
\cos(\alpha,\beta) & = & \frac{4z^2 - 2(|a|x + |b|y + 
|c|z)}{\delta_{\alpha}(p) \delta_{\beta}(p)} \nonumber \\
\label{CosAngle} & & \\
   \cos(\beta,\mu) & = & \frac{4x^2 + 4z^2 +|b|^2 - 
|c|^2}{\delta_{\beta}(p) \delta_{\mu}(p)} \nonumber
\end{eqnarray}

For the f\/irst conf\/iguration one has $\delta_{\alpha}(p) = 
d_1$, $\delta_{\beta}(p) = d_2$ and $\cos(\alpha,\beta) = 
\cos\theta$, thus yielding the equations
\begin{eqnarray}
                  y^2 + z^2 & = & \frac{1}{4} (d_1^2 - |a|^2) 
\nonumber \\ 
\label{EqFirstG6}
        4x^2 + (2z - |c|)^2 & = & d_2^2 - |b|^2 \\
4z^2 -2(|a|x + |b|y + |c|z) & = & d_1d_2\cos\theta \; . \nonumber
\end{eqnarray}
This is a system of three quadratic equations with six unknowns, 
the three coordinates $(x,y,z)$ of the point $p$, and the three 
coordinates $(|a|,|b|,|c|)$ in the parameter space of the 
$\sixthG$ manifold (see \cite{GR02}). An algebraic solution of 
these equations for $(x,y,z)$ in terms of $(|a|,|b|,|c|)$, or 
vice versa, would in general yield higher degree (decoupled) 
equations  for each variable, and thus are not so illuminating. 
Particular solutions can be obtained by (i) assuming specif\/ic 
values for the parameters $(|a|,|b|,|c|)$, and then calculating 
numerically the position of $C_0$, or (ii) assuming some 
particular position for $C_0$, and then calculating the 
parameters $(|a|,|b|,|c|)$. This second method does not follow 
the strategy of determining the parameters of the manifold using 
two more triples of clusters of galaxies (see Section~\ref{Decide}), 
thus it will not be pursued here. The next section presents
examples of application of the f\/irst method.

Finally, let us examine the second conf\/iguration which is
simpler. One has $\delta_{\beta}(p)=d_1$, $\delta_{\mu}(p)=d_2$
and $\cos(\beta,\mu) = \cos\theta$, thus yielding the equations
\begin{eqnarray}
       4x^2 + (2z - |c|)^2 & = & d_1^2 - |b|^2 \nonumber \\
\label{EqSecondG6}
       4x^2 + (2z + |c|)^2 & = & d_2^2 - |b|^2 \\
4x^2 + 4z^2 + |b|^2 -|c|^2 & = & d_1d_2\cos\theta \; . \nonumber 
\end{eqnarray}
These equations can be partially solved giving
\begin{eqnarray}
        z & = & \frac{1}{8|c|} \, (d_2^2 - d_1^2) \nonumber \\
\label{ModelSecondG6}
      |c| & = & \frac{1}{2} \sqrt{d_1^2 + d_2^2 - 
2d_1d_2\cos\theta} \\ 
x^2 + z^2 & = & \frac{1}{16} \, (d_1^2 + d_2^2 + 2
d_1d_2\cos\theta) - \frac{1}{4} \, |b|^2 \; . \nonumber
\end{eqnarray}
In this case the $y-\,$coordinate is not constrained by the 
conf\/iguration of topological images, since the only orthogonal 
transformation involved in the calculations has the $y-\,$axis 
as its axis of rotation.

\section{Numerical Examples}
\label{Examples}

Let us now apply the results obtained in the previous section to the proposed
multiple images of Roukema and Edge \cite{RE97}, in a FL universe whose matter
components are pressureless dust and a cosmological constant. The models
presented below are small universes with compactif\/ication scales much
smaller than the horizon radius, so they may seem to be in conf\/lict with
constraints on the topology coming from observations of the CMBR. However, it
must be recalled that all current constraints for f\/lat universes hold
exclusively for models with (i) toroidal spatial sections \cite{Torus,Inoue},
or (ii) any f\/lat (compact and orientable) spatial section, but in
cosmological models without a dark energy component, and moreover, with
the observer located on the axis of rotation of a screw motion of the
corresponding covering group \cite{FlatCMB}. As has been shown by Inoue
\cite{Inoue}, the addition of a cosmological constant term makes the
constraints less stringent, whereas the ef\/fect of considering the observer
out of an axis of rotation is totally unknown. Since the models presented
below consider both, a cosmological constant term, and the observer of\/f
an axis of rotation, they can not be considered as being ruled out by current
observational data.

The models constructed here consider $C_1$ as being the cluster RXJ
1347.5--1145 and $C_2$ the cluster CL 09104+4109. Then for the values
$\Omega_{m0} = 0.3$ and $\Omega_{\Lambda 0} = 0.7$, one has $d_1 = 1158h^{-1}
\, Mpc$, $d_2 = 1142h^{-1} \, Mpc$ and $\theta = 87^o$. Other examples can be
built by simply reversing these identif\/ications, i.e. by considering $C_1$
as being the cluster CL 09104+4109 and $C_2$ the cluster RXJ 1347.5--1145. As
before, let us f\/irst examine the classes $\secondG-\fifthG$. One has $|b| =
1142h^{-1} \, Mpc$, and because of the expression $\sqrt{d_1^2\sin^2 \theta -
|a|^2}$ in Table~\ref{Tb:Positions} one also has the constraint
\begin{equation}
\label{|a|-Max}
|a| \leq 1156.4h^{-1} \, Mpc \; .
\end{equation}
The models within class $\secondG$ are special because they 
have a f\/ixed value of $y$, say $y = - 30.3h^{-1} \, Mpc$; 
however the $z-\,$coordinate depends on the parameter $|a|$, 
and remarkably is sensible to this value as can be seen with 
the following two examples.
\begin{enumerate}
\item First consider the case when $|a|$ is slightly lower than 
the maximum value allowed by (\ref{|a|-Max}), say $|a|=1156h^{-1} 
\, Mpc$. Then $z = \pm 15.5h^{-1} \, Mpc$.
\item Second consider the symmetric case when $|a| = |b| = 
1142h^{-1} \, Mpc$. In this case one has $z = \pm \, 91.0h^{-1} 
\, Mpc$.
\end{enumerate}

\begin{table}[t]
\begin{center}
\begin{tabular}{*{6}{|c}|} \hline & & \multicolumn{2}{c|}{} & 
\multicolumn{2}{c|}{} \\
Class & $|a| \, (h^{-1} \, Mpc)$ & \multicolumn{2}{c|}{$y \; (h^{-1} 
\, Mpc)$}  & \multicolumn{2}{c|}{$z \; (h^{-1} \, Mpc)$}  \\ 
& & \multicolumn{2}{c|}{} & \multicolumn{2}{c|}{} \\ \hline 
\hline 
$\thirdG$ & $1156$ & $-21.4$ & $-39.2$ & $-32.9$ & $-2.0$ \\ 
\cline{2-6}
          & $1142$ &  $22.2$ &  $82.8$ & $108.5$ & $73.5$ \\ 
\hline
$\fourthG$ & $1156$ & $-14.9$ &  $-45.8$ &  $-45.8$ & $-14.9$ \\ 
\cline{2-6}
           & $1142$ &  $60.7$ & $-121.3$ & $-121.3$ & $60.7$ \\ 
\hline
$\fifthG$ & $1156$ &  $-3.5$ &  $-57.1$ & $-67.9$ & $-37.0$ \\ 
\cline{2-6}
          & $1142$ & $127.3$ & $-187.9$ & $143.5$ & $38.5$ \\ 
\hline
\end{tabular}
\caption[Examples of models within classes $\thirdG-\fifthG$.] 
{\label{Tb:Examples} \footnotesize Examples of models within 
classes $\thirdG-\fifthG$.}
\end{center}
\end{table}

Note that the classes $\thirdG-\fifthG$ do not yield models with
a f\/ixed value of $y$; instead, both $y$ and $z$ depend on the
parameter $|a|$. In Table~\ref{Tb:Examples} we show the values 
of $y$ and $z$ calculated from Table~\ref{Tb:Positions} for $|a| 
= 1156$ and $1142h^{-1} \, Mpc$. In this table the f\/irst column 
for each coordinate corresponds to the f\/irst solution of 
Table~\ref{Tb:Positions}, and the second column for the second 
solution.

Now we deal with models within class $\sixthG$. For the f\/irst 
conf\/iguration one obtains from eqs. (\ref{EqFirstG6})
\[
(2x + |a|)^2 + (2y + |b|)^2 = d_1^2 + d_2^2 - 2d_1d_2\cos\theta 
- |c|^2 \; ,
\]
which implies that 
\[
|c|^2 \leq d_1^2 + d_2^2 - 2d_1d_2\cos\theta \; .
\]
Furthermore, from the f\/irst and third equations in 
(\ref{EqFirstG6}) one also has
\[
|a| \leq d_1 \qquad\mbox{and}\qquad |b| \leq d_2 \; .
\]
A family of simple examples are obtained by taking $|a| = d_1$. 
In fact, in this case one has
\[
y=z=0 \qquad , \qquad x = - \frac{1}{2} d_2\cos\theta \qquad , 
\qquad |b|^2 + |c|^2 = d_2^2\sin^2\theta \; .
\]
Thus, taking $|b|=|c|$, one model of a universe with spatial 
sections of class $\sixthG$ that f\/its the f\/irst
conf\/iguration with the Roukema--Edge hypothesis is
\[
|a| = 1158h^{-1} \, Mpc \qquad\mbox{and}\qquad |b| = |c| = 
1140.4h^{-1} \, Mpc \; ,
\]
with Coma located at
\[
x = - 29.9h^{-1} \, Mpc \qquad\mbox{and}\qquad y=z=0 \; .
\]

On the other side, for the second conf\/iguration one has
\begin{equation}
\label{ExampleSecondG6}
|c| = 791.6h^{-1} \, Mpc \quad\mbox{,}\quad z = -5.8h^{-1} \, 
Mpc \quad\mbox{,}\quad |b| \leq 834.1h^{-1} \, Mpc \; ,
\end{equation}
the last inequality being obtained from the last equation in 
(\ref{ModelSecondG6}). It is illustrative to give two specif\/ic 
examples as done with the $\secondG$ models.
\begin{enumerate}
\item First consider the case when $|b|$ is slightly lower than 
its maximum value allowed by (\ref{ExampleSecondG6}), say $|b| = 
834h^{-1} \, Mpc$, then one has $x = \pm 7h^{-1} \, Mpc$.
\item Second consider the symmetric case when $|b| = |c| = 
791.6h^{-1} \, Mpc$. In this case $x = \pm 131.5h^{-1} \, Mpc$.
\end{enumerate}

\section{The case of three triples of images}  
\label{Decide}

In this section it is shown that the discovery of two 
additional triples of clusters of galaxies close to 
$\{C_0,C_1,C_2\}$ would allow the determination of the 
topology of the universe, and in most cases the determination 
of its size. Let us denote by $\{D_0,D_1,D_2\}$ and 
$\{E_0,E_1,E_2\}$ these two additional triples of topological 
images. Mathematically, to characterize the \emph{closeness} 
relation between two triples $\{C_i\}$ 
and $\{D_i\}$ it suf\/f\/ices the lengths of the geodesic 
segments $\overline{C_iD_i}$ ($i=0,1,2$) to be the same and 
smaller than the injectivity radius. Observationally, it is 
enough that $C_0$ and $D_0$ are two nearby clusters of 
galaxies, while the distances between $C_i$ and $D_i$ 
($i=1,2$) are equal (within the observational error bounds) 
to the distance between $C_0$ and $D_0$.%
\footnote{Strictly speaking, this \emph{closeness} relation 
is not a necessary condition, but observationally it would 
be simpler to look for other triples of images in the 
neighborhood of the f\/irst one.}

By parallel transporting the triangle $C_1D_1E_1$ along the 
geodesic segment $\overline{C_0C_1}$, one obtains two triangles 
with a common vertex, namely the triangle of nearby clusters 
$C_0D_0E_0$, and that of \emph{transported} clusters of 
$C_1D_1E_1$. It is just a matter of elementary analytic geometry 
to determine the unique rotation that takes one triangle to the 
other. Note however that one can easily f\/ind also the unique 
ref\/lection that takes one triangle to the other, if it exists. 
If the angle of rotation is dif\/ferent from $\pi$, $2 \pi/3$, 
$\pi/2$ or $\pi/3$, then the isometry that takes $C_0$ to $C_1$ 
is not a screw motion, but a ref\/lection, and the Universe would 
be spatially non-orientable. On the contrary, if the angle of 
rotation is either $\pi$, $2 \pi/3$, $\pi/2$ or $\pi/3$, then one 
can think this is not by coincidence, so the Universe would be 
spatially orientable. In such a case, if the angle of rotation is
dif\/ferent from $\pi$, it uniquely determines to which class the 
topology of the Universe belongs, namely $\thirdG$, $\fourthG$ or 
$\fifthG$ respectively.%
\footnote{Note however that if there exists a ref\/lection 
taking one triangle to the other, in order to settle def\/initely 
the orientability of space, it would be necessary to identify a 
fourth triple of multiple images.}

Let us restrict our analysis to the orientable case in order to 
be specif\/ic. The determination of the rotation taking 
$C_0D_0E_0$ to the parallel transportation of $C_1D_1E_1$ 
provides also the direction of the axis of rotation of the screw 
motion linking $C_0$ with $C_1$. If the Universe has a topology 
of class $\thirdG$, $\fourthG$ or $\fifthG$, the translation 
vector is parallel to this axis, so elementary geometry can be 
used to determine the parameter $|a|$ and the position of the 
axis. Moreover, the isometry linking $C_0$ with $C_2$ has to be 
a translation, and a parallel transport of the triangle 
$C_2,D_2,E_2$ to $C_0D_0E_0$ would conf\/irm it. A remarkable 
fact is that, if the topology of the Universe has been 
identif\/ied to be of class $\thirdG$, $\fourthG$ or $\fifthG$, 
the vector $c$ is automatically f\/ixed, and observational 
searches can be performed to f\/ind the topological images of 
$C_0$, $D_0$ and $E_0$ due to the isometries $\gamma$ and 
$\gamma^{-1}$ for validation of the model.

Let us now consider the case when the angle of rotation taking 
$C_0D_0E_0$ to the parallel transport of $C_1D_1E_1$ is $\pi$. 
In this case the topology of the Universe has to be of class 
$\secondG$ or $\sixthG$. One can decide between these two 
possibilities by parallel transporting $C_2D_2E_2$ to 
$C_0D_0E_0$. If the angle of rotation between these triangles 
is null, then the isometry linking $C_0$ to $C_2$ is a 
translation, and the Universe has topology of class $\secondG$. 
On the other hand, if the angle of rotation is $\pi$, the 
Universe has topology of class $\sixthG$. In the former case 
one can proceed as before and determine the length $|a|$ and 
the position of the axis of the screw motion. However, since 
for the class $\secondG$, the vector $c$ is a free parameter, 
its modulus and direction remain undetermined.

If the topology of the Universe turns out to be of class 
$\sixthG$, the multiple images can be f\/itted within the two 
inequivalent conf\/igurations described in 
Section~\ref{ModBuild}. One can decide between both 
conf\/igurations by just looking at the directions  of the 
axes of rotation, for if they are orthogonal the f\/irst 
conf\/iguration would be the correct one, while if they are 
parallel the correct one is the second. Using elementary 
geometry one can completely determine the three axes of 
rotation and translations (thus determining the global 
shape of space) if the multiple images f\/it with the 
f\/irst conf\/iguration. On the other hand, with the second 
conf\/iguration one can determine the vectors $b$ and $c$, 
and thus the direction of $a$, but it is impossible to 
determine the length $|a|$, as could have been anticipated 
from eqs.~(\ref{EqSecondG6}). However, in this latter case, 
one can design ef\/fective search procedures to look for 
multiple images due to the isometries $\alpha$ and 
$\alpha^{-1}$, thus providing at least robust constraints 
for the parameter $|a|$ (see~\cite{BerGo}).

\section{Discussion and Further Remarks}
\label{Concl}

The work presented in this paper originated with the
following problem in the context of cosmological models
with f\/lat spatial sections: Suppose we have identif\/ied
three clusters of galaxies as being dif\/ferent topological
images of the same object. How do these multiple images
constrain the possible models for the shape of our Universe?
A natural extension of this work would be the study of this
problem in the context of universes with non--f\/lat spatial
sections, specif\/ically those with positive curvature (since
multiply connected spaces of negative curvature are very
unlikely to have a detectable topology \cite{Detect}).

It has been shown here that one can accommodate any of the
six classes of compact orientable 3--dimensional f\/lat space
forms to f\/it with any conf\/iguration of three topological
images of a cosmic object. It can be seen from the
construction of the models that one could also easily f\/it
any of the non--orientable f\/lat manifolds. Moreover, the
main result in this paper is that the identif\/ication of
two more triples of multiple images of clusters of galaxies,
in the neighbourhood of the f\/irst one, is enough to
completely determine the topology of space, as well as
its size in most of the cases.

Even if the primary goal of this paper is not to construct specif\/ic
candidates for the shape of our universe, but to present a systematic
procedure for building such models, it turns out that the illustrative
examples constructed by using the Roukema--Edge hypothesis are not in
contradiction with current observational data.

In view of these results, it seems of primary importance to state and test
hypotheses like that of Roukema and Edge, i.e. that the clusters RXJ
1347.5--1145 and CL 09104+4109 are topological images of the Coma cluster,
since the identif\/ication of a very small quantity of multiple images is,
as has been shown here, enough to determine (or almost determine) the global
shape of the universe. The problem of testing this kind of hypothesis can be
solved by the Local Noise Correlations (LNC) method proposed in \cite{BerGo}.
The problem of generating such kind of hypotheses seems to be much harder,
although current ef\/forts are being done to f\/ind multiple images of our
Galaxy \cite{Galaxy}, clusters of galaxies \cite{Rou} and radio-loud AGNs
\cite{RMBS}.

To close this paper, let us stress that there have only been 
considered here models in which the topological images are related
by the generators of the covering groups of the corresponding 
manifolds. This needs not be the case, for one could also 
consider other isometries (compositions of the generators) as 
being the responsible for the multiple images. Thus the list 
of possible models presented here is not exhaustive.

\vspace{3mm}
\section*{Acknowledgments}

I would like to thank CLAF/CNPq and FAPESP (contract 02/12328-6)
for the grants under which this work was carried out, and to the
CBPF for kind hospitality. I also thank to Bruno Mota, Marcelo
Rebou\c{c}as and Armando Bernui for critical reading of previous
versions of this work and for their valuable suggestions.


%
%
\end{document}